\documentclass[12pt,a4paper]{amsart}
\usepackage{amssymb,amsmath,amsthm}
\usepackage{graphicx,float}
\usepackage{hyperref}
 \usepackage[color=green!40]{todonotes}
 \usepackage{soul}

\theoremstyle{plain}
\newtheorem{thm}{Theorem}[section]
\newtheorem{theorem}{Theorem}[section]
\newtheorem{lemma}[theorem]{Lemma}

\newtheorem{proposition}[theorem]{Proposition}

\theoremstyle{definition}

\newtheorem{claim}[theorem]{Claim}

\newcommand{\opposite}{\textit{opposite}}
\newcommand{\same}{\textit{same}} 

\title{Finding the diameter of a tree with distance queries}
\date{}

\begin{document}

\author[]{D\'aniel Gerbner$^a$, Andr\'as Imolay$^{b}$, Kartal Nagy$^b$,\\ Bal\'azs Patk\'os$^a$, Krist\'of Z\'olomy$^{b}$  
\\
\small $^a$ HUN-REN Alfr\'ed R\'enyi Institute of Mathematics \\
\small $^b$ ELTE E\"otv\"os Lor\'and University 
}
\begin{abstract}
    We study the number of distance queries needed to identify certain properties of a hidden tree $T$ on $n$ vertices. A distance query consists of two vertices $x,y$, and the answer is the distance of $x$ and  $y$ in  $T$. We determine the number of queries an optimal adaptive algorithm needs to find two vertices of maximal distance up to an additive constant, and the number of queries needed to identify the hidden tree asymptotically. 
We also study the non-adaptive versions of these problems, determining the number of queries needed exactly.

\end{abstract}

\maketitle

\section{Introduction}

Learning a hidden subgraph or some properties of a hidden subgraph is a well-studied problem in search theory, with applications among others in DNA sequencing \cite{A+,DFS}, phylogenetics \cite{philo1,philo2}. We are given vertices $v_1,\dots,v_n$, and we know that there is a hidden graph $G$ on these vertices. We are also given some types of queries that we can ask, and some property of $G$ that we want to determine. The question is how many queries are needed in the worst case.

Typical queries are the following. We can ask two vertices $u,v$ and the answer tells whether $uv$ is an edge of $G$ \cite{AN,RS}. Or we can ask a set $U$ of vertices and the answer tells whether there is any edge inside $U$ \cite{AA,AC}. Some of the typical goals are completely determining (reconstructing) $G$, or determining $G$ up to isomorphism, or finding out whether $G$ is connected \cite{ACK}. The most amount of research was done on the reconstruction problem \cite{os} of certain graph classes: different types of random graphs \cite{B,KM}, bounded degree graphs \cite{SW}, and some further graph classes \cite{BG,BG2}.

Here we deal with \textit{distance queries} (also called shortest path queries). As a query, we can ask two vertices $v_i,v_j$, and the answer is the distance $d(v_i,v_j)$ of $v_i$ and $v_j$, i.e., the length of the shortest path in $G$ between $v_i$ and $v_j$.

Our starting point is a Mathoverflow question by Curious (Raphael Coxteau) \cite{stack}. We are given a hidden tree $T$ on $n$ vertices, one can ask distance queries and the goal is to find two vertices with the largest distance in $T$. Our goal is to determine the smallest number $f_A(n)$ of queries such that $f_A(n)$ queries are always enough to find two such vertices. An upper bound $f_A(n)\le 2n-4$ is given in \cite{stack}. Let us describe the algorithm. We start by picking a vertex $u$ and asking the queries $u,u'$ for each vertex other than $u$. Then we pick an arbitrary vertex $v$ that is at the largest distance from $u$, and query $v$ with each vertex $v'$ except for $u$ and an arbitrary neighbor $u_0$ of $u$. Observe that one of the neighbors of $u$ are at distance $d(u,v)-1$ from $v$, while the others are at distance $d(u,v)+1$. Therefore, we can figure out $d(v,u_0)$, thus we know the distance of $v$ from any vertex. The largest of these distances is the largest distance in $T$.

For the sake of completeness, we present the proof of the correctness of this algorithm.
We need to show that there is a longest path that contains $v$. Let us recall that for any two vertices $x,y$ there is a unique path between $x$ and $y$ in $T$, we call it the \textit{$x-y$ path}. Assume that the $x-y$ path is a longest path. Let $z$ be the closest vertex to $u$ on the $x-y$ path. If the $u-v$ path shares at least one vertex with the $x-y$ path, then they share $z$. Then the $z-v$ path is at least as long as the $z-x$ path, thus the $y-v$ path is at least as long as the $y-x$ path. If the $u-v$ path does not share any vertex with the $x-y$-path, then the $u-v$ path contains a vertex $z'$ closest to $z$ on the $u-z$ path. Then the $z'-x$ path goes through $z$ and is not longer than the $z'-v$ path. Therefore, the
path from $v$ to $z'$, then to $z$, then to $y$ is not shorter than the $x-y$ path.

The trivial lower bound $f_A(n)\ge n-1$ is provided in \cite{stack}, along with some ideas on how it may be improved to roughly $n+\log n$. Our main theorem is the following bound.

\begin{theorem}\label{main}
    For any $n$ we have $f_A(n)\ge 2n-9$.
\end{theorem}

We also consider the non-adaptive version of this problem, where all the queries have to be asked at the same time, thus the information provided by earlier answers cannot be used. We denote the number of queries needed to find two vertices at maximum distance by $f_N(n)$. We determine the exact value of $f_N(n)$ for large enough $n$.

\begin{proposition}\label{nonaddia}
    For every $n \geq 13$ we have $f_N(n)=\frac{n(n-3)}{2}$.
\end{proposition}

Let us mention the variant of this problem where the goal is to determine the diameter $d$ instead of finding two vertices at distance $d$. Another variant is when we want to find both $d$ and two vertices at distance $d$. The number of queries needed may differ somewhat for these three problems (for example if $n=3$, then no queries are needed to show that the diameter is $2$). However, the number of queries needed to solve these variants differ only in a small additive constant, hence we lose little generality by only considering our original problem. In fact, in the non-adaptive case and $n\ge 13$,
the number of queries needed is still $\frac{n(n-3)}{2}$, where the upper bound follows from Proposition \ref{nonadiso} and the lower bound follows from Lemma \ref{lemi}.

We also consider two related problems, determining the hidden tree $T$ completely, and up to isomorphism. The minimum number of queries needed to determine $T$ is denoted by $g_A(n)$ in the adaptive case and $g_N(n)$ in the non-adaptive case. The minimum number of queries needed to determine $T$ up to isomorphism is denoted by $h_A(n)$ in the adaptive case and $h_N(n)$ in the non-adaptive case.

We remark that all these problems are trivial for general graphs $G$ (or even for general connected graphs) instead of trees. The upper bound $\binom{n}{2}$ holds because by asking every pair as a query, we learn each edge, since the corresponding pairs are answered $1$. In the case the answer is $1$ to each of the first $\binom{n}{2}-1$ queries, $G$ may be $K_n$, but it is possible that the last pair $u,v$ are formed by non-adjacent vertices. In this case, we do not know the diameter, nor $G$ up to isomorphism. We know that $u$ and $v$ are at maximum distance, but it is easy to see that after $\binom{n}{2}-2$ queries are answered $1$, we cannot show two vertices at maximum distance.

Determining the hidden graph has been studied for several graph classes, including trees.
In particular, Reysin and Srivastava \cite{RS} gave the bound $g_A(n)=\Omega(n^2)$. However, there are two small problems with their proof. First, they refer to a paper \cite{alon} that proves a quadratic lower bound in the non-adaptive case, and apply it as if it was an adaptive result. Second, their proof is based on the \textit{spider}, which is a tree with vertices $u$, $v_1,\dots, v_{\lfloor n/2\rfloor}$ and $w_1,\dots, w_{n-\lfloor n/2\rfloor-1}$ such that $u$ is joined to $v_i$ and $v_i$ is joined to $w_i$ for each $1 \leq i \leq \lfloor n/2\rfloor$. If $u$ is known, the only meaningful queries are of the form $v_i,w_j$, and the answer tells whether there is an edge between $v_i,w_j$. In this case, the problem is equivalent to finding a given matching on $n-1$ vertices. However, we also have a bipartition of the vertices to $v_i$ and $w_j$, and Reysin and Srivastava \cite{RS} do not argue why this cannot give any useful information. 

We will prove the quadratic bound directly, without using \cite{alon}, and also determine the asymptotics.

\begin{proposition}\label{prop:adrec}
   $g_A(n)=\frac{n^2}{4}+O(n)$.
\end{proposition}

We show that even if the hidden tree is given up to isomorphism, we may still need quadratic many queries to determine it.

\begin{proposition}\label{spid}
Let $T$ be a spider, i.e., a tree with vertices $u$, $v_1,\dots, v_{\lfloor n/2\rfloor}$ and $w_1,\dots, w_{n-\lfloor n/2\rfloor-1}$ such that $u$ is joined to $v_i$ and $v_i$ is joined to $w_i$ for each $1 \leq i \leq \lfloor n/2\rfloor$.
        The smallest number of queries in an adaptive algorithm that determines a hidden copy of $T$ is $\frac{n^2}{8}+O(n)$.
\end{proposition}

In the case we want to determine the tree up to isomorphism, we still need quadratic many queries.
Here, our best known lower and upper bound differ by a factor of $2$. We conjecture that neither the lower nor the upper bound is sharp.

\begin{proposition}\label{adiso}
   $\frac{n^2}{8}-O(n)\le h_A(n)\le\frac{n^2}{4}+O(n)$.
\end{proposition}

Finally, we completely resolve the remaining non-adaptive problems.

\begin{proposition}\label{nonaddet}
  For every $n\ge 5$ we have $g_N(n)=\lceil\frac{n(n-2)}{2}\rceil$.
\end{proposition}

\begin{proposition}\label{nonadiso}
    For every $n \geq 13$ we have $h_N(n)=\frac{n(n-3)}{2}$. 
\end{proposition}

\subsection{Notation} 

As it is usual in Search Theory, we look at our problems as games between two players, the Questioner and the Adversary. There is no tree given at the beginning, Questioner asks queries and Adversary gives the answers such that they are consistent with our assumption that the graph is a tree. The game ends when the goal of the algorithm is reached, i.e., there are two vertices at maximum distance in all the trees consistent with the answers in the case of $f_A(n)$, all the trees consistent with the answers are isomorphic in the case of $h_A(n)$, and there is only one tree consistent with the answers in the case of $g_A(n)$. The goal of Questioner is to end the algorithm as fast as possible, and the goal of the Adversary is the opposite. Clearly, if both players play optimally, then the game is finished after exactly $f_A(n)$ (resp. $g_A(n)$ and $h_A(n)$) queries. Indeed, if Questioner follows the optimal algorithm, then the game is finished after at most $f_A(n)$ queries, while if an algorithm follows Questioner's strategy, then it is finished no later than the game.

We will use standard graph theoretic terminology. In particular, given a graph $G$, we denote its vertex set by $V(G)$ and its edge set by $E(G)$. Trees are often denoted by $T$. Given sets $U$ and $V$ of vertices, $G[U]$ denotes the graph with vertex set $U$ and edge set that consists of the edges of $G$ that are inside $U$. $G[U,V]$ denotes the graph with vertex set $U\cup V$ and edge set that consists of the edges of $G$ with one endpoint in $U$ and the other in $V$. It is well-known that a graph can be partitioned to maximal connected subgraphs, called \textit{connected components}. Here we will refer to the vertex sets of connected components as \textit{components}.

\begin{figure}
    \centering
\begin{center}
\begin{tikzpicture}[scale=1, every node/.style={draw, circle, fill=black, inner sep=1.5pt}]

    \node[fill=red, label=above: $u$] (A) at (-1,0) {};
    \node[label=above: $v$] (B) at (1,0) {};
    
    \draw (A) -- (B);
    
    \foreach \i in {1,2} {
        \node[fill=blue] (A\i) at (-2, \i-2.5) {};
        \draw (A) -- (A\i);
    }
    \foreach \i in {3,4} {
        \node[fill=red] (A\i) at (-2, \i-2.5) {};
        \draw (A) -- (A\i);
    }
    
    \foreach \i in {1,2,3,5} {
        \node (B\i) at (2, \i-3) {};
        \draw (B) -- (B\i);
    }
    \foreach \i in {4} {
        \node[fill=red, label=above:$v'$] (B\i) at (2, \i-3) {};
        \draw (B) -- (B\i);
    }

\end{tikzpicture} \hspace{ 1cm}
\begin{tikzpicture}[scale=1, every node/.style={draw, circle, fill=black, inner sep=1.5pt}]

    \node[fill=red, label=above: $u$] (A) at (-1,0) {};
    \node[label=above: $v$] (B) at (1,0) {};
    
    \draw (A) -- (B);
    
    \foreach \i in {3,4} {
        \node[fill=red] (A\i) at (-2, \i-2.5) {};
        \draw (A) -- (A\i);
    }
    
    \foreach \i in {1,2,3,5} {
        \node (B\i) at (2, \i-3) {};
        \draw (B) -- (B\i);
    }
    \foreach \i in {4} {
        \node[fill=red, label=above:$v'$] (B\i) at (2, \i-3) {};
        \draw (B) -- (B\i);
    }

    \node[fill=blue] (C1) at (3,0.5) {};
    \node[fill=blue] (C2) at (3,1.5) {};
    \draw (B4) -- (C1);
    \draw (B4) -- (C2);
\end{tikzpicture}
    \caption{A double star and a real caterpillar, whose partial distance graph only differs on edges between blue and red vertices.}
    \label{fig:doublestar-caterpillar}
\end{center}
\end{figure}
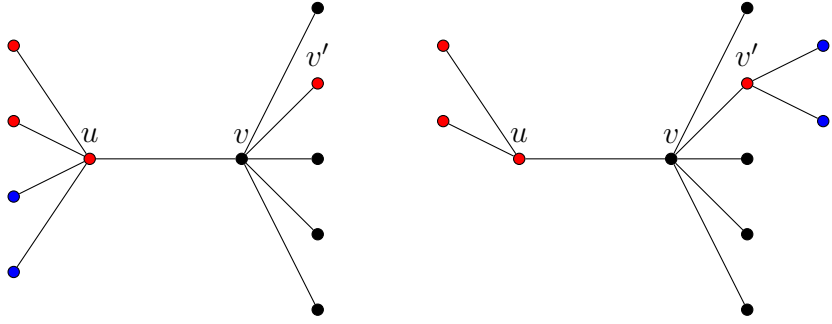

We will consider the following specific trees. A \textit{double star} is a tree with two non-leaf vertices $u$ and $v$. In other words, a double star has an edge $uv$ and $n-2$ additional vertices joined to either $u$ or $v$, such that at least one and at most $n-3$ of the additional vertices are joined to $u$. We say that a tree is a \textit{real caterpillar} if it has exactly 3 vertices of degree at least 2. These 3 vertices must induce a path in the tree. Note that the diameter of real caterpillars is 4. For an example, see Figure \ref{fig:doublestar-caterpillar}.

In the non-adaptive case, the queries form the \textit{query graph}. A \textit{distance graph} of a graph $G$ is a graph on $V(G)$ where the edges have integer weights and for each edge $uv$ of the distance graph the weight is the distance of $u$ and $v$ in $G$. We call a tree $T$ \emph{consistent}, if the query graph $Q$ weighted by the answers is a distance graph of $Q$.

\section{Proofs} 

\subsection{Adaptive strategies} \

First, we prove our main theorem, Theorem~\ref{main}. After some definitions, we start with three technical lemmas.

Let $T$ be a tree and $G$ a weighted graph on the same vertex set with a positive integer weight $w(e)$ on each edge $e$. We say that $G$ is a \textit{partial distance graph} of $T$, if the weight of every edge $xy \in E(G)$ is exactly the distance of $x$ and $y$ in $T$.

Let $D$ be a double star on the vertex set $U \cup V$, where $U \cap V=\emptyset$, with $u \in U$ and $v \in V$ being the centers of the double star, and the edges of $D$ are $uv$ and all the pairs of the form $uu'$ and $vv'$ where $u' \in U \setminus \{u\}$ and $v' \in V \setminus \{v\}$. We call $U$ and $V$ the \emph{sides} of the double star. Note that the centers and the sides uniquely determine the double star.

\begin{lemma}\label{pdg2}
    Let $D$ be a double star with sides $U$ and $V$ and centers $u \in U$ and $v \in V$ and let $G$ be a partial distance graph of $D$. Assume that $C \subsetneq U \setminus \{u\}$ is a connected component of $G[U]$, and there exists a vertex $v' \in V \setminus \{v\}$ such that there is no edge joining $v'$ and a vertex of $C$ in $G$. Then $G$ is a partial distance graph of a real caterpillar.
\end{lemma}

\begin{proof}
    Let $D'$ be the real caterpillar obtained from $D$ by removing the edges joining $u$ to vertices of $C$ and adding the edges joining $v'$ to vertices of $C$. Clearly, $D'$ is a real caterpillar. Furthermore, $G$ is a partial distance graph of $D'$ as well, because
    the only distances changed are between the sets $(U \setminus C) \cup \{v'\}$ and $C$, and such vertex pairs are not edges of $G$ by the conditions of the lemma. 
\end{proof}

\begin{lemma}\label{edges2}
    Let $G$ be an $n$-vertex graph with $V(G)$ being the disjoint union of $X$ and $Y$ where $2\le |X|\le |Y|$. Let $x\in X$ and $y\in Y$ be fixed such that every $x'\in X\setminus \{x\}$ is connected to at least one vertex in every component of $G[Y]$ except for the one containing $y$, and every  $y'\in Y\setminus \{y\}$ is connected to at least one vertex of every component of $G[X]$ except for the one containing $x$. Furthermore, assume that there exists $x'\in X \setminus \{x\}$ and a subgraph $F \subseteq G[X,Y]$ such that $d_F(x')\le 1$ and $|E(F)|\ge n-1$. Then $|E(G)|\ge 2n-7$. 
\end{lemma}

\begin{proof}
    Let $c_X$ and $c_Y$ denote the number of connected components of $G[X]$ and $G[Y]$. If $c_X\ge 5$, then by the assumption all vertices in $Y\setminus \{y\}$ have degree at least 4 in $G[X,Y]$ and so $|E(G)|\ge 4|Y|-4\ge 4\frac{n}{2}-4=2n-4$. If $c_X\le 4$, then $G[X]$ contains at least $|X|-4$ edges, and $G[Y]$ contains at least $|Y|-c_Y$ edges. Finally, $G[X,Y]$ contains at least $|E(F)|+c_Y-2$ edges, where the extra comes from the number of edges incident to $x'$ but not in $F$. Altogether, $G$ contains at least $|X|-4+|Y|-c_Y+|E(F)|+c_Y-2\ge 2n-7$ edges.
\end{proof}

\begin{lemma}\label{three_components}
Let $G$ be a bipartite graph with 3 connected components, $C$, $C'$ and $C^{''}$. Assume that $C$ has at least two vertices. Then we can add two edges to $G$ in a way that the obtained graph $G'$ is a connected bipartite graph containing a spanning tree with a leaf $w \in C' \cup C^{''}$ such that if the partite sets of $G'$ have different size, then $w$ is in the smaller partite set of $G'$.
\end{lemma}

\begin{proof}
    Let $C_1$, $C_2$ be the partite sets of $C$ with $|C_1| \leq |C_2|$. Similarly, let $C'_1$, $C'_2$ be the partite sets of $C'$ with $|C'_1| \leq |C'_2|$ and let $C^{''}_1$, $C^{''}_2$ be the partite sets of $C^{''}$ with $|C^{''}_1| \leq |C^{''}_2|$. Note that by the conditions, $C_1$, $C_2$, $C'_2$ and $C^{''}_2$ are nonempty, and $C'_1$ (resp. $C^{''}_1$) is empty if and only if $C'$ (resp. $C''$) is an isolated vertex.
    
    Let $F$, $F'$ and $F^{''}$ be spanning trees of $C$, $C'$ and $C^{''}$, respectively. We prove that we can add two edges $e$ and $f$ such that $E_0:=E(F) \cup E(F') \cup E(F^{''}) \cup \{e,f\}$ is the edge set of a spanning tree of $G'=G \cup \{e,f\}$ having a leaf in $C' \cup C^{''}$ which is in the smaller partite set of $G'$ (if there is a smaller partite set).

    \textit{Case 1:} $F'$ has a leaf in $C'_1$ or $F^{''}$ has a leaf in $C^{''}_1$.

    By symmetry, we can assume that $F'$ has a leaf $w$ in $C'_1$. Let $x \in C_1$, $y \in C'_2$ and $z \in C^{''}_3$, and let $e=xy$, $f=xz$. Then one of the partite sets of $G'=G \cup \{e,f\}$ is $C_1 \cup C'_1 \cup C^{''}_1$, which cannot be larger than the other partite set, and $w$ is a leaf of $E_0$. Note that this also proves the lemma if $|C'_1|=|C'_2|$ or $|C^{''}_1|=|C^{''}_2|$ by changing the role of $C'_1$ and $C'_2$ or $C^{''}_1$ and $C^{''}_2$. Hence from now we can assume that $|C'_1| < |C'_2|$ and $|C^{''}_1| <|C^{''}_2|$.

    \textit{Case 2:} There is no leaf of $F'$ in $C'_1$ and no leaf of $F^{''}$ in $C^{''}_1$.

    We can assume that $|C'_2|-|C'_1| \geq |C^{''}_2|-|C^{''}_1|$. 

    \textit{Case 2.1:} $C^{''}_2$ contains at least two vertices.

    Then $F^{''}$ has at least one edge, hence it has a leaf $x$, which must be in $C^{''}_2$. Let $x' \in C^{''}_2$ be a vertex other than $x$. Choose vertices $y \in C_2$, $z \in C'_2$ and $t \in C_1$ arbitrarily, and add the edges $e=x'y$ and $f=zt$ to $G$. Now $G'=G \cup \{e,f\}$ is connected, $x$ is a leaf of $E_0$ and the partite set $C_1 \cup C'_1 \cup C^{''}_2$ containing it is not larger than the other partite set of $G'$.
    
    \textit{Case 2.2:} $C^{''}_2$ is an isolated vertex $x$.

    This means that $C^{''}_1$ is empty, hence $C^{''}$ is the isolated vertex $x$. This case is similar to Case~2.1, but now $F^{''}$ has no edges, hence we will make $x$ to be a leaf by adding an edge incident to it.
    Let $y \in C_2$, $z \in C'_2$ and $t \in C_1$ be arbitrary vertices, and add the edges $e=xy$ and $f=zt$. Now $G'=G \cup \{e,f\}$ is connected, $x$ is a leaf of $E_0$ and the partite set $C_1 \cup C'_1 \cup C^{''}_2$ containing it is not larger than the other partite set of $G'$. The proof is finished.
\end{proof}

Now we are ready to prove Theorem~\ref{main}, which states that at least $2n-9$ queries are needed to find two vertices at maximum distance.

\begin{proof}[Proof of Theorem \ref{main}]
    We are going to describe a possible strategy of the Adversary. Roughly speaking, he will always reply in a way that his answers form a partial distance graph of a double star, and as long as the conditions of Lemma~\ref{pdg2} hold, the Questioner cannot distinguish between a double star with diameter 3, and a real caterpillar with diameter 4. Adversary's aim is that once this is not true anymore, the graph obtained by the queries should almost satisfy the conditions of Lemma \ref{edges2}. 

    Whatever the first query $uv$ is, the Adversary answers 1 and declares the additional information that the graph he is hiding is either a double star $D$ with centers $u$ and $v$ or a real caterpillar. In particular, the Adversary guarantees that his answers are consistent with a double star $D$ with centers $u$ and $v$. In a double star $D$ with sides $U$ and $V$ and centers $u \in U$, $v \in V$, if the centers are given, then the distance of any two vertices $w$ and $z$ determines whether they are on the same side or not, and conversely, if we know whether $w$ and $z$ are on the same side, we know their distance. Hence, we get an equivalent problem after the first query and the guarantees of the Adversary, if we consider \emph{side queries} instead of distance queries, i.e., for a query of two vertices the Adversary tells the Questioner whether they are on the same side in the double star or not. If they are on the same side, the Adversary answers \emph{same}, otherwise he answers \emph{opposite}.
    
    At each point throughout the process, the Adversary maintains two auxiliary graphs, $G_1$ and $G_2$, with vertex sets being equal to the vertex set of the hidden tree. Until a certain point, $G_1$ always consists of the edges corresponding to the queries with answer \opposite.
    Similarly, the edges of $G_2$ correspond to the queries with answer \same. In particular, after the first query, $uv$ is the only edge of $G_1$ and $G_2$ has no edge. We never remove edges from $G_1$ and $G_2$, however, at a certain point we will add two additional edges to $G_1$, which can be considered as additional information given by the Adversary. The idea is to add these additional edges according to Lemma~\ref{three_components}, which will guarantee that after the Questioner can prove that the hidden tree cannot be a real caterpillar, the conditions of Lemma~\ref{edges2} are satisfied by $G_1 \cup G_2$. For an illustration, see Figure \ref{fig:mainthm}.

    \begin{figure}
        \centering
         \begin{minipage}{0.3\textwidth}
     \centering
    \begin{tikzpicture}[scale=1, every node/.style={draw, circle, fill=black, inner sep=1.5pt}]
        \node (u) at (0,5) [label=left:$u$] {};
        \node (v) at (1,5) [label=right:$v$] {};
        \draw[color=red] (u) -- (v);

        \foreach \i in {1,2,3,4}{
        \node (b\i) at (0,5- 0.8*\i) [label=left: \scriptsize $\ell_\i$]{};
        \node (j\i) at (1,5 -0.8*\i) [label=right: \scriptsize $r_\i$]{};
        }
        \node (b5) at (0,5- 0.8*5) [label=left: \scriptsize $\ell_5$]{};
        \node (b6) at (0,5- 0.8*6) [label=left: \scriptsize $\ell_6$]{};
        \draw[color=red] (b5) -- (j3);
        \draw[color=red] (b6) -- (j4);
        \draw[color=blue] (b5) -- (b6);

        \node[color=red] (j3') at (1,5-0.8*3) {};
        \node[color=blue] (b1') at (0,5-0.8*1) {};
        \node[color=blue] (b3') at (0,5-0.8*3) {};
        \node[color=blue] (b4') at (0,5-0.8*4) {};

        \draw[color=red] (b1) -- (j1);

        \draw[color=red] (b2) -- (j2);

        \draw[color=red] (b2) -- (j1);

        \draw[color=blue] (j1) -- (j2);

        \draw[color=red] (b3) -- (j1);
        \draw[color=red] (b2) -- (j3);
        \draw[color=red] (b3) -- (j4);
        \draw[color=red] (b4) -- (j4);
        \draw[color=red] (u) -- (j1);
        \draw[color=red] (b2) -- (j1);
        \draw[color=red] (u) -- (j2);

        \draw[color=blue] (b1) to[out=180, in=180] (b3);
        \draw[color=blue] (u) to[out=180, in=180] (b2);
        \draw[color=blue] (b3) -- (b4);

        \draw[color=blue] (v) -- (j1);
        \draw[color=blue] (j2) to[out=0, in=0] (j4);
    \end{tikzpicture}
     \end{minipage}
     \begin{minipage}{0.5\textwidth}
         \begin{center}
    \begin{tikzpicture}[scale=0.8, every node/.style={draw, circle, fill=black, inner sep=1.4pt}]
            \node (u) at (0,-0.5) [label=above:\scriptsize$u$] {};
            \node (v) at (1,-0.5) [label=above:\scriptsize$v$] {};
            \draw(u) -- (v);
    
            \foreach \i in {1,2,3,4}{
            \node (b\i) at (-1,2-0.667*\i) [label=left: \scriptsize $\ell_\i$]{};
            \node (j\i) at (2,2 - \i) [label=right: \scriptsize $r_\i$]{};
            \draw (u) -- (b\i);
            \draw (v) -- (j\i);
            }
            \node (b5) at (-1,2-0.667*5) [label=left: \scriptsize $\ell_5$]{};
            \node (b6) at (-1,2-0.667*6) [label=left: \scriptsize $\ell_6$]{};
            \draw (u) -- (b5);
            \draw (u) -- (b6);

    \end{tikzpicture} 
    
    \vspace{5 mm}
    
    \begin{tikzpicture}[scale=0.8, every node/.style={draw, circle, fill=black, inner sep=1.5pt}]
            \node[draw=none,fill=none] at (-3,0){};
            \node (u) at (0,-0.5) [label=above:\scriptsize$u$] {};
            \node (v) at (1,-0.5) [label=above:\scriptsize$v$] {};
            \draw(u) -- (v);
    
            \foreach \i in {2,5,6}{
            \node (b\i) at (-1,2-0.667*\i) [label=left: \scriptsize $\ell_\i$]{};
            \draw (u) -- (b\i);
            }
            \foreach \i in {1,2,3,4}{
            \node (j\i) at (2,2 - \i) [label=above: \scriptsize $r_\i$]{};
            \draw (v) -- (j\i);
            }
            \foreach \i in {1}{
            \node (b\i) at (3,0-0.5*\i) [label=right: \scriptsize $\ell_\i$]{};
            \draw (j3) -- (b\i);
            }
            \foreach \i in {3,4}{
            \node (b\i) at (3,0.5-0.5*\i) [label=right: \scriptsize $\ell_\i$]{};
            \draw (j3) -- (b\i);
            }
    \end{tikzpicture}
    \end{center}
     \end{minipage}
        \caption{The graph on the left represents a query graph at a point of the process, with red edges being part of $G_1$, and blue edges forming $G_2$. The condition of Lemma \ref{edges2} is violated by vertex $r_3$ and the component $\{l_1,l_3,l_4\}$, so both graphs on the right side are consistent with queries made so far.}
        \label{fig:mainthm}
    \end{figure}
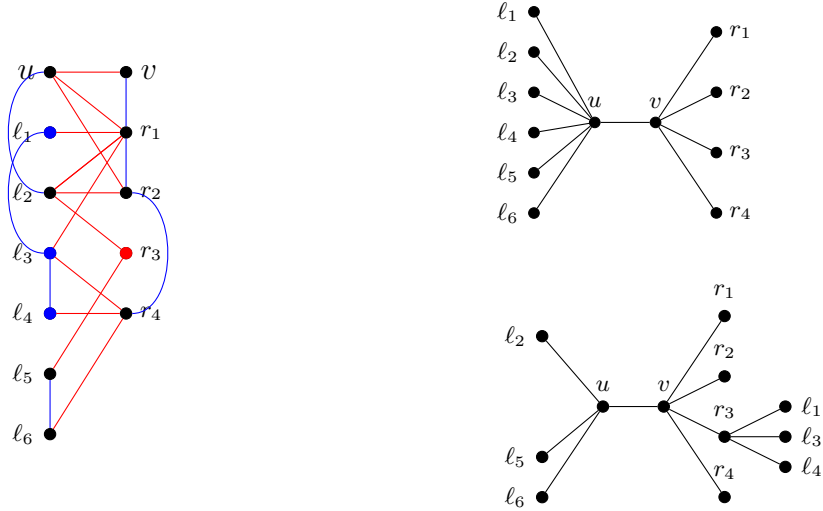

    Now we are ready to define the strategy of the Adversary precisely. For any query $wz$, the Adversary always answers ``opposite'' except if it is not possible, i.e., if $w$ and $z$ are in the same partite set of the same connected component of $G_1$. After the answer, we add the edge to $G_1$ if the Adversary answered ``opposite'', and add it to $G_2$ if the answer was ``same''. 

    There is one \emph{special moment} in the strategy, the first time when $G_1$ has exactly three components. At this point, apply Lemma~\ref{three_components} to $G_1$ and its components, with $C$ being the component containing $u$ and $v$. Add the two additional edges $e$ and $f$ given by Lemma~\ref{three_components} to $G_1$. Furthermore, we assume that the information given by these edges indeed holds for the hidden graph, and Adversary reveals this to the Questioner, i.e., the Adversary gives the information that if the Questioner asked one of the queries corresponding to $e$ or $f$ at this point or a later point of the algorithm, then the answer would be ``opposite''. 

    We finished describing the strategy. At any point of the process, we say that a tree is \emph{consistent}, if it satisfies the answers and the additional information given by the Adversary. Call the first time when a real caterpillar is not consistent the \emph{end of the process}. We aim to show that at the end of the process, $G_1 \cup G_2$ satisfies the conditions of Lemma~\ref{edges2}.
    Observe that at every point of the process, $G_1$ is a bipartite graph, and every edge of $G_2$ connects vertices in the same connected component of $G_1$. It follows that the connected components of $G_1 \cup G_2$ are always the same as the connected components of $G_1$. First, we show that the process cannot end before the special moment, i.e., if $G_1 \cup G_2$ has at least three components, then there exists a consistent real caterpillar.

    Consider a point of the process before the special moment, after the first query, and let $C'$ and $C^{''}$ be two components of $G_1 \cup G_2$ not containing $u$ and $v$. Let $K$ be one of the components of $G_2$ contained in $C'$, and $x \in C^{''}$. Note that $K$ is contained in one of the partite sets of $C'$, hence all of the vertices of $K$ are on the same side in every consistent double star. There is a consistent double star $D$ such that $x$ and $K$ are on different sides, because in any consistent double star, we can swap the partite sets of $C'$ between the sides to get another consistent double star. Hence the conditions of Lemma~\ref{pdg2} hold with $D$, $v'=x$, and connected component $K$. Consequently, there is a real caterpillar consistent with the answers.

    Finally, we prove that the conditions of Lemma~\ref{edges2} hold at the end of the process. After the special moment, 
    $G_1$ is a connected bipartite graph with partite sets $U$ and $V$, with $u\in U, v\in V$. Note that the partite sets do not change from this point.
    If there exists an $x' \in U \setminus \{u\}$ that is not connected to a component of $(G_1 \cup G_2)[V]=G_2[V]$ not containing $v$, then the conditions of Lemma~\ref{pdg2} hold with $x'$ and this component, which contradicts to the fact the there is no consistent real caterpillar. Similarly, every $y' \in V \setminus \{v\}$ is connected to each component of $(G_1 \cup G_2)[U]$ not containing $u$.
    Furthermore, Lemma~\ref{three_components} guarantees that $G_1$ contains a spanning tree $F$ with a leaf different from $u$ and $v$ in one of the partite sets, which is not larger than the other partite set. Hence the conditions of Lemma~\ref{edges2} are indeed satisfied, proving that $G_1 \cup G_2$ has at least $2n-7$ edges. All edges of $G_1 \cup G_2$ correspond to a query, apart from the two edges added at the special moment, hence $2n-9$ queries are necessary.
\end{proof}

Let us continue with the proof of Proposition \ref{prop:adrec}, which states that $n^2/4+O(n)$ queries are needed to determine the hidden tree.

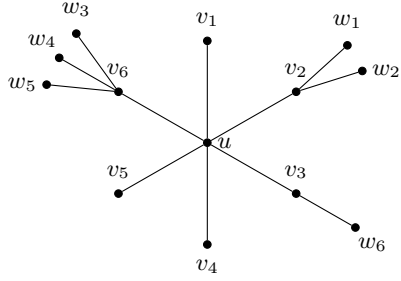
\begin{figure}
    \centering
    \begin{tikzpicture}[scale=1.5, every node/.style={draw, circle, fill=black, inner sep=1pt}]
    
        \node[label=right:\scriptsize $u$] (C) at (0,0) {};
    
        \foreach \i in {1,2,3,4,5,6} {
            \node (L\i) at (90-\i*60:0.9) {};
            \draw (C) -- (L\i);
        }
        \foreach \k in {1,2,3,5,6} {
         \node[draw=none, fill=none] at ({90-(\k-1)*60}:0.9) [label=above: \scriptsize $v_{\k}$] {};
        }
         \node[draw=none, fill=none] at ({90-(4-1)*60}:0.9) [label=below: \scriptsize $v_{4}$] {};        
        \node (A1) at (90+5-60:1.5) [label=above: \scriptsize $w_1$]{};
        \node (A2) at (90-5-60:1.5) [label=right: \scriptsize $w_2$]{};
        \draw (L1) -- (A1);
        \draw (L1) -- (A2);
    
        \node (B1) at (90-60*2:1.5) [label=south east: \scriptsize $w_6$]{};
        \draw (L2) -- (B1);
    
        \node (E1) at (90-10-60*5:1.5)[label=above: \scriptsize $w_3$] {};
        \node (E2) at (90-60*5:1.5) [label=north west: \scriptsize $w_4$] {};
        \node (E3) at (90+10-60*5:1.5) [label=left: \scriptsize $w_5$]{};
        \draw (L5) -- (E1);
        \draw (L5) -- (E2);
        \draw (L5) -- (E3);
    \end{tikzpicture}
    \caption{An example for a hidden graph in the proof of Proposition \ref{prop:adrec}.}
    \label{fig:adapt-determ-lower}
\end{figure}

\begin{proof}[Proof of Proposition \ref{prop:adrec}]
    For the lower bound, let us restrict ourselves to graphs consisting of a vertex $u$, $\lfloor n/2\rfloor$ neighbors of $u$, $v_1,\dots, v_{\lfloor n/2\rfloor}$, and $n-\lfloor n/2\rfloor-1$ leaves $w_i$, each connected to a neighbor of $u$. For an example, see Figure \ref{fig:adapt-determ-lower}. The Adversary tells the partition of the vertices to these three sets for free. Then the answers are fully determined except for queries between $v_i$ and $w_j$, where the answer may be 1 or 3, and between $w_i$ and $w_j$, where the answer may be 2 or 4. In both cases, he answers the larger number if such an answer is possible, i.e, it is still possible after the answer that the hidden graph is a tree satisfying the extra properties we started with. We can assume that the Questioner believes the extra information provided by the Adversary, thus does not ask queries whose answers are already known. In particular, no query is asked where the answer is 1 or 2.

    Assume that we identified the hidden tree $T$. Let $W_i$ denote the set of leaf neighbors of $v_i$ for $1 \leq i \leq \lfloor n/2\rfloor$. Let $G_{ij}$ denote the graph with vertex set $\{v_i,v_j\}\cup W_i\cup W_j$, and two vertex is connected if we made a query between them. Note that we assumed that no query was asked with answer 1 or 2, hence $\{v_i\}\cup W_i$ spans no edges for all $1 \leq i \leq \lfloor n/2\rfloor$. It follows that the graphs $G_{ij}$ ($1 \leq i \leq \lfloor n/2\rfloor$) cover each query exactly once, hence we need to bound the sum of the number of edges in these graphs. If $G_{ij}$ has a connected component $C$ that contains neither $v_i$ nor $v_j$, then we could swap $C$: connect its vertices in $W_i$ to $v_j$ and its vertices in $W_j$ to $v_i$. This would be consistent with all the answers, which contradicts with the fact that we identified $T$. It follows that $G_{ij}$ has at most two connected components, so it has at least $|W_i|+|W_j|$ edges. Consequently, the number of queries asked is at least
    \[ \sum_{1 \leq i<j \leq \lfloor n/2\rfloor} |W_i|+|W_j|=(\lfloor n/2\rfloor-1) \sum_{i=1}^{\lfloor n/2\rfloor} |W_i|= (\lfloor n/2\rfloor-1)(n-\lfloor n/2\rfloor-1).\]

 \smallskip

    For the upper bound, we pick an arbitrary vertex $u$ and query it with all the other vertices. Then we obtain levels of the tree such that each edge is between consecutive levels. We ask each pair between consecutive levels, except between $u$ and the first level. Since this is a bipartite graph, there are at most $(n-1)^2/4$ such queries. We find the hidden tree, since for each edge of the tree, the two endpoints were queried, so we determine all the edges.
\end{proof}

Let us continue with Proposition \ref{spid}, which states that $\frac{n^2}{8}+O(n)$ queries are needed to determine a hidden spider.

\begin{proof}[Proof of Proposition \ref{spid}]
    For the lower bound, we assume first for simplicity that $n$ is odd. The Adversary tells the partition of the vertices to $\{u\}$, $\{v_1 , v_2,  \ldots , v_{\lfloor n/2 \rfloor}\}$, $\{w_1 , w_2, \ldots , w_{\lfloor n/2 \rfloor}\}$. Then the answers are determined, except for queries between $v_i$ and $w_j$, where the answer can be 1 or 3. The Adversary answers to such queries 3 if possible, i.e., he answers 1 only if each of the trees isomorphic to $T$ and consistent with the answers contains the edge $v_iw_j$. Therefore, such a query gives no additional information, hence we can assume that no such query was asked, thus no query was answered 1.

    Assume that we have determined the hidden copy of $T$ where $v_i$ is connected to $w_{\sigma(i)}$ for some permutation $\sigma$ of $1,\dots,(n-1)/2$. We claim that for each $i$ and $j$, at least one of the pairs $v_i,w_{\sigma(j)}$ and $v_j,w_{\sigma(i)}$ was queried. Otherwise, since $v_iw_{\sigma(i)}$ and $v_jw_{\sigma(j)}$ were not queried, the hidden copy of $T$ could have the edges $v_iw_{\sigma(j)}$ and $v_jw_{\sigma(i)}$ instead, a contradiction. Therefore, we have asked at least $\binom{(n-1)/2}{2}$ queries.
    In the case $n$ is even, the same holds for the $\binom{(n-2)/2}{2}$ pairs $i,j$ where $v_i$ and $v_j$ are not leaves, giving the lower bound $\binom{(n-2)/2}{2}$.

 \smallskip

    For the upper bound, we assume that $n\ge 7$. We want to find $u$ first. We pick a vertex $x$ and query it with all the other vertices. If $n$ is odd, the number of queries answered 1 tells us whether $x$ is $u$, $v_i$ for some $i$, or $w_j$ for some $j$, since those vertices have different degrees in $T$.
    If $n$ is even, then $v_{n/2}$ has degree 1, like a vertex $w_j$, but has more than one vertices at distance 2, hence in this case we can also identify whether $x$ is $u$, $v_{n/2}$, $v_i$ for some $i<n_2$ or $w_j$ for some $j$. If $x$ is $u$, we found $u$. If $x$ is $w_j$, then $u$ is the only vertex at distance 2 from $x$. If $x$ is $v_{n/2}$, $u$ is the only neighbor of $x$. If $x$ is $v_i$, we query a neighbor $y$ of $x$ with a non-neighbor of $x$. If the answer is 1 or 2, then $y$ is $u$, otherwise $y=w_i$ and the other neighbor of $x$ is $u$.

    Having $u$, we want to partition the vertices to neighbors and non-neighbors of $u$. We can do this by querying $u$ with all the other vertices (in fact, it is not needed, the distance from $x$ already gives this information). Now we pick a neighbor $v_1$ of $u$, and query it with all the non-neighbors of $u$ to find the other neighbor $w_1$ of $v_1$. 
    Then we pick another neighbor $v_2$ of $u$, and find the other neighbor of $v_2$. This time we query it with all the non-neighbors of $u$, except for $w_1$. We continue this way, each time one fewer query is needed, thus we used $\binom{n-\lfloor n/2\rfloor}{2}$ queries after finding the partition. Earlier we used $O(n)$ queries, completing the proof.
\end{proof}

Finally, we deal with Proposition \ref{adiso}, which states that at least $n^2/8-O(n)$ queries are needed to determine the hidden tree up to isomorphism. The proposition also states an upper bound $n^2/4+O(n)$, but it is an immediate corollary of Proposition \ref{prop:adrec}.

\begin{proof}[Proof of Proposition \ref{adiso}] Let $n\ge 6$.
    The Adversary decides that $T$ is a spider, but does not tell this to the Questioner
    and answers as in the proof of Proposition \ref{spid}. Now the queries that are answered 1 according to these rules are not useless, since we have to prove that the hidden tree is indeed a spider. Assume that we have determined $T$ up to isomorphism. Then $T$ is a spider with vertices $u,v_1,\dots, v_{\lfloor n/2\rfloor},w_1,\dots,w_{\lceil n/2\rceil -1}$ such that $u$ is joned to $v_i$ and $v_i$ is joined to $w_i$ for each $1\le i\le \lfloor n/2\rfloor$. Assume that for some $i$, $v_i,w_i$ was not queried. If there is a $j\neq i$ such that $w_i$ was not queried with $v_j$, nor with $w_j$, then the answers are consistent with the tree where $w_i$ is adjacent to $v_j$ instead of $v_i$. That tree has two vertices of degree more than $2$ since $n\ge 6$, thus not isomorphic to $T$, a contradiction. If $v_i,w_i$ was queried, and later $v_j,w_j$ was also queried, they were both answered 1. 
    Then we claim that at least one of the pairs $v_i,w_j$ and $v_j,w_i$ was queried before $v_i,w_i$. Indeed, otherwise it would have been possible to answer 3 to the query $v_i,w_i$, since the spider obtained by replacing the edges $v_iw_i$ and $v_jw_j$ by the edges $v_iw_j$ and $v_jw_i$ is consistent with these answers.

    We obtained that no matter whether $v_i,w_i$ and $v_j,w_j$ were queried, there was a query taking one element of $\{v_i,w_i\}$ and one element of $\{v_j,w_j\}$. There are $n^2/8-O(n)$ pairs $i,j$, completing the proof.
\end{proof}

\subsection{Non-adaptive strategies}

Let us start with the proof of Proposition \ref{nonaddet}. Recall that it states that for $n\ge 5$, $\lceil \frac{n(n-2)}{2}\rceil$ queries are needed and enough to determine a hidden tree in the non-adaptive setting.

\begin{proof}[Proof of Proposition \ref{nonaddet}]
For the lower bound, observe that if a query graph determines the hidden tree, then every vertex has degree at least $n-2$. Indeed, if $uv$ and $uw$ are not queried, it is possible that the hidden tree has a vertex $x$ of degree $n-2$, joined to each vertex but $u$, and $u$ is joined to either $v$ or $w$. 

For the upper bound, let the query graph be a complete graph with a matching of size $\lfloor n/2\rfloor$ removed. We claim that we learn for every pair $u,v$ whether there is an edge joining them. If the pair was queried, the answer obviously gives this information. 
Observe that if $u$ and $v$ are adjacent, then for any other vertex $w$, $d(u,w)$ and $d(v,w)$ differ by exactly one. 
If $u$ and $v$ are not adjacent, then this is possible only if there is a path $uxyv$ and $u$ and $v$ are leaves. 
But if the answers are consistent with such a tree, then we know that $x$ is the only neighbor of $u$, besides potentially $v$, and $y$ is the only neighbor of $v$, besides potentially $u$. If $x,y$ was queried, we know whether they are neighbors, in which case $uv$ is not an edge, or not, in which case $uv$ is an edge. Therefore, $x,y$ was not queried and we can say the same things about the pair $(x,y)$ instead of $(u,v)$. In particular, the neighbors of $x$ are $u$ and potentially $y$, and the neighbors of $y$ are $v$ and potentially $x$. But this means that no edge leaves the set $\{u,v,x,y\}$, thus $T$ is disconnected as $n\ge 5$, a contradiction.    
\end{proof}

Finally, we will prove Propositions \ref{nonaddia} and \ref{nonadiso} together.

\begin{thm}\label{nemad}
    The number of queries needed in a non-adaptive algorithm to find a pair of vertices at maximum distance, or to determine the hidden tree up to isomorphism is $\frac{n(n-3)}{2}$ if $n \geq 13$. Moreover, a query graph $Q$ finds a pair of vertices at maximum distance if and only if $Q$ determines the hidden tree up to isomorphism, which holds if and only if every vertex has degree at least $n-3$.
\end{thm}

The next lemma immediately implies the lower bound.

\begin{lemma}\label{lemi}
    If a non-adaptive query graph $Q$ determines the diameter or finds two vertices with maximum distance, then the minimum degree of $Q$ is at least $n-3$.
\end{lemma}

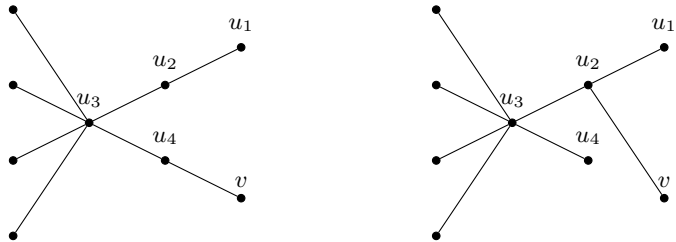
\begin{figure}
    \centering
    \begin{tikzpicture}[scale=1, every node/.style={draw, circle, fill=black, inner sep=1pt}]
    
        \node[label=above:\scriptsize$u_3$] (A) at (-1,0) {};
        
        \draw (A);
        
        \foreach \i in {1,2,3,4} {
            \node (A\i) at (-2, \i-2.5) {};
            \draw (A) -- (A\i);
        }
    
        \node (C) at (0,0.5) [label=above:\scriptsize$u_2$] {};
        \draw (C);
        \draw (A) -- (C);
        \node (D) at (0,-0.5) [label=above:\scriptsize$u_4$] {};
        \draw (D);
        \draw (A) -- (D);
    
        \node (E) at (1,1) [label=above:\scriptsize$u_1$] {};
        \draw (E);
        \draw (C) -- (E);
        \node (F) at (1,-1) [label=above:\scriptsize$v$] {};
        \draw (F);
        \draw (D) -- (F);

    \end{tikzpicture}
    \hspace{2 cm}
    \begin{tikzpicture}[scale=1, every node/.style={draw, circle, fill=black, inner sep=1pt}]
        \node[label=above:\scriptsize$u_3$] (A) at (-1,0) {};
        
        \draw (A);
        
        \foreach \i in {1,2,3,4} {
            \node (A\i) at (-2, \i-2.5) {};
            \draw (A) -- (A\i);
        }
    
        \node (C) at (0,0.5)[label=above:\scriptsize$u_2$] {};
        \draw (C);
        \draw (A) -- (C);
        \node (D) at (0,-0.5) [label=above:\scriptsize$u_4$] {};
        \draw (D);
        \draw (A) -- (D);
    
        \node (E) at (1,1) [label=above:\scriptsize$u_1$]{};
        \draw (E);
        \draw (C) -- (E);
        \node (F) at (1,-1) [label=above:\scriptsize$v$] {};
        \draw (F);
        \draw (C) -- (F);
    
    \end{tikzpicture} 
    \caption{Two trees for which only distances $u_1v, u_2v, u_4v$ differ.}
    \label{fig:nonad-diam-lower}
\end{figure}

\begin{proof}
    Let $Q$ be such a query graph and assume $v$ is the vertex with degree at most $n-4$. Then consider the following two trees $T,T'$ on the same vertex set (see Figure \ref{fig:nonad-diam-lower} for illustration). Let $T$ consist of a path $u_1u_2u_3u_4v$ on five vertices and $n-5$ pendant edges from $u_3$ and let $T'$ be the graph obtained from $T$ by removing the edge $u_4v$ and adding the edge $u_2v$. Observe first, that the diameter of $T$ is 4, while the diameter of $T'$ is 3. As $v$ is a leaf in both graphs, only distances involving $v$ can change and actually only 3 of them do, the distances to $u_1,u_2,u_4$. Therefore, if an Adversary arranges the vertices $u_1,u_2,u_4$ to be the vertices of $Q$ with $(u_1v),(u_2v),(u_4v)\notin E(Q)$ and answers queries $Q$ according to the distances of $T$ and $T'$, then it is impossible to distinguish $T$ and $T'$ while these two trees have different diameters. Moreover, the only pair with maximum distance in $T$ is $u_1,v$, while their distance in $T'$ is $2$, thus smaller than the diameter. This completes the proof.
\end{proof}

We turn to prove the upper bound. From now on, we assume that a query graph $Q$ is given with minimum degree at least $n-3$, and we received the answers $q(x,y)$ to all of the queried pairs $x$ and $y$. We call a tree $T$ \emph{consistent}, if the graph $Q$ weighted by the answers is a distance graph of $T$. We start with several lemmas.

\begin{lemma} \label{null}
    For any four distinct vertices $a,b,c,d$, there exists a vertex $v$ such that $v$ was queried with all of them.
\end{lemma}

\begin{proof}
    For each vertex among $a,b,c,d$, there are at most two vertices not connected to it in the query graph, hence there are at most $8$ vertices not connected to at least one vertex among $a,b,c,d$. Consequently, as $n \geq 13$, there is a vertex $v$ connected to $a$, $b$, $c$ and $d$.
\end{proof}

\begin{lemma} \label{parity}
    For any two vertices $a$ and $b$, their distance in any two consistent trees has the same parity.
\end{lemma}

\begin{proof}
    Let $v$ be a vertex queried with both $a$ and $b$, such vertex exists by Lemma~\ref{null}. In any consistent tree, the distance of $a$ and $b$ modulo 2 is equal to $q(a,v)+q(b,v)$ modulo 2.
\end{proof}

\begin{lemma}\label{lemm1}
    There are no consistent hidden trees $T_0$ and $T_1$ such that $a,b,c,d$ are on a path in this order in $T_0$ (but they are not necessarily consecutive vertices), while these vertices are also in a path in $T_1$, in the order of $b,a,d,c$ or $c,a,d,b$ (but not necessarily consecutively).
\end{lemma}

\begin{proof}
    Let $v$ be a vertex queried with all of $a,b,c,d$, such vertex exists by Lemma~\ref{null}. Consider the maximum number among the obtained answers $q(v,a), q(v,b), q(v,c), q(v,d)$. As $T_0$ is consistent, this maximum cannot be $q(v,b)$ or $q(v,c)$, and as $T_1$ is consistent, it also cannot be $q(v,a)$ or $q(v,d)$, which is a contradiction.
\end{proof}

\begin{lemma}\label{lemm2}
    There are no consistent hidden trees $T_0$ and $T_1$ such that $a,b,c,d$ are on a path in this order in $T_0$, while in $T_1$, $'bad'$ is a path and $c$ has distance 2 from $a$, but $c$ is not a neighbor of $b$, nor $d$.
\end{lemma}

\begin{proof}
    Let $v$ be a vertex queried with all of $a,b,c,d$, such vertex exists by Lemma~\ref{null}. Consider the maximum number among the obtained answers $q(v,a), q(v,b), q(v,c), q(v,d)$. As $T_0$ is consistent, this maximum cannot be $q(v,b)$ or $q(v,c)$. Similarly, as $T_1$ is consistent, it is either $q(v,c)$ or $q(v,b)=q(v,d)$. These observations together imply a contradiction. 
\end{proof}

\begin{lemma}\label{lemnew1}
    If there are hidden trees $T_0,T_1$ consistent with the answers such that $abcd$ is a path in $T_0$ and $dabc$ is a path in $T_1$, then for every vertex $v$, the distance of $b$ and $v$ is the same in $T_0$ and $T_1$.
\end{lemma}

\begin{proof} 
Observe that $d$ was not queried with $a$ and $c$, thus it was queried with all $v\not\in \{a,b,c,d\}$. For $a$ (and similarly for $c$), there is at most one vertex $v \notin\{a,b,c,d\}$ such that $a$ and $v$ were not queried. 

Deleting $a,b,c,d$ cuts $T_i$ into several trees ($i=0,1$). Each of those trees has exactly one vertex joined to exactly one of $a,b,c,d$. We denote by $F_i(a)$ the sets of vertices in those trees that have a vertex joined to $a$. We define $F_i(b),F_i(c),F_i(d)$ analogously. 

If a vertex $v\not\in \{a,b,c,d\}$ was queried with both $a$ and $c$, consider the smallest value among $q(v,a),q(v,c),q(v,d)$. First, assume that this smallest value is unique. If it is $q(v,a)$, then $q(v,d)-q(v,a)$ takes different values in $T_0$ and $T_1$, a contradiction. Similarly, if $q(v,c)$ is the minimum, then $q(v,d)-q(v,c)$ takes different values in $T_0$ and $T_1$, a contradiction. Finally, if $q(v,d)$ is the minimum, then $q(v,a)-q(v,d)$ takes different values in $T_0$ and $T_1$, a contradiction again. Hence the smallest value among $q(v,a),q(v,c),q(v,d)$ is not unique. It follows that $v\in F_0(b)\cap F_1(b)$.
    
If a vertex $v$ is in $F_i(b)$, then its distance from $b$ in $T_i$ is $q(v,d)-2$, thus the statement of the lemma is true for all $v \in F_0(b)\cap F_1(b)$. 

If there is a vertex $v\not\in \{a,b,c,d\}$ not in $F_i(b)$ for some $i$, then $v$ was not queried with either $a$ or $c$, in particular there are at most two such vertices. If $F_i(d)$ is not empty for some $i$, then we know a vertex $d'\in F_i(d)$ that is a neighbor of $d$ since every vertex not in $ \{a,b,c,d\}$ was queried with $d$. In this case, we also know that $d' \in F_{1-i}(d)$.
 Observe that $d'$ was not queried with $a$ nor $c$, since their distance is different in $T_0$ and $T_1$, thus in this case there are no other vertices that are not in $F_i(b)$ for some $i$, and the statement of the lemma holds as $d'b$ has distance 3 both in $T_0$ and $T_1$.
 
 We consider now the case that $F_0(d)$ and $F_1(d)$ are both empty.
 If $F_0(c)$ is not empty, then there is a vertex $c'\in F_0(c)$ that is a neighbor of $c$ in $T_0$. Then $q(c',d)=2$, hence $c'\in F_1(a)$, and $c'$ is a neighbor of $a$ in $T_1$. In particular, $c'$ was not queried with $a$ nor $c$. It follows that there are no other vertices that are not in $F_0(b) \cap F_1(b)$, and the statement of the lemma holds as $c'b$ have distance 2 both in $T_0$ and $T_1$. A similar argument deals with the case when $F_1(a)$ is nonempty. 

Finally, we consider the case when $F_0(c)$ and $F_1(a)$ are also empty. 
If $F_1(c)$ is not empty, then let $v\in F_1(c)$. The distance of $v$ and $b$ in $T_1$ is $q(v,d)-2$. As $F_0(c)$ and $F_0(d)$ are empty, we know that $v$ is in $F_0(a)$ or $F_0(b)$. It follows that the distance of $v$ and $b$ in $T_0$ is also $q(v,d)-2$, as wanted. A similar argument for vertices in $F_0(a)$ finishes the proof of the lemma.
\end{proof} 

\begin{lemma}\label{lemnew2}
    For every pair of consistent trees $T_0, T_1$, there is a vertex $z$ such that the distance of $z$ and $v$ is the same in $T_0$ and $T_1$ for every vertex $v$.
\end{lemma}

\begin{proof}
    Assume that $u$ and $v$ are vertices such that we cannot decide whether they are adjacent or not, i.e., we have a consistent tree $T_0$ where $u$ and $v$ are not adjacent, and a consistent tree $T_1$ where $uv$ is an edge. Observe that $u,v$ was not queried and there is at most one other vertex $u'$ such that $u,u'$ was not queried, and at most one other vertex $v'$ such that $v,v'$ was not queried. We denote the distance function in $T_0$ and $T_1$ with $d_0$ and $d_1$, respectively. In particular, if $x$ and $y$ was queried, then $d_0(x,y)=d_1(x,y)=q(x,y)$. 

Let $uw_1w_2\dots w_iv$ be the path between $u$ and $v$ in $T_0$. By Lemma~\ref{parity}, $i$ is even. If $i\ge 4$, then $d_0(v,w_1)\ge d_0(u,w_1)+3$, while $|d_1(v,w)-d_1(u,w)|=1$ for every vertex $w$. This shows that $w_1$ is either $u'$ or $v'$, and analogously the same holds for $w_i$. If $i\ge 6$, then the same holds for $w_2$, a contradiction, thus $i\le 4$. If $i=4$, we know that $q(u,w_2)=q(v,w_3)=2$ and $q(u,w_3)=q(v,w_2)=3$. In $T_1$, $w_2$ is closer to $u$ than to $v$ and $w_3$ is closer to $v$ than to $u$, thus the shortest path between them is through the edge $uv$. This is a contradiction by Lemma~\ref{lemm1}. Hence $i=2$ and $d_0(u,v)=3$. Therefore, by letting $x=w_1$ and $y=w_2$, the hidden tree $T_0$ contains the path $uxyv$. Our next goal is to show that either $vuxy$ or $xyvu$ is a path in $T_1$.

\textsc{Case I}  We queried $u$ with $x$. Because of $T_0$, we have $q(x,u)=1$, thus $vux$ is a path in $T_1$. We know that $y$ was queried with at least one of $x,u,v$. 

If $x,y$ was queried, then $q(x,y)=1$, hence $vuxy$ is a path in $T_1$, as we wanted. 

If $u,y$ was queried, then $q(u,y)=2$. There are three options. Either $vuxy$ is a path, as wanted, or $yvux$ is a path, which is not possible by Lemma~\ref{lemm1}, or $y$ is not a neighbor of $v$ and $x$, which is not possible by Lemma~\ref{lemm2}.

Finally, if $v,y$ was queried, then $q(v,y)=1$, implying that $yvux$ is a path in $T_1$, contradicting Lemma~\ref{lemm1}.

\textsc{Case II}  We did not query $u$ and $x$. Note that we can also assume that we did not query $v$ with $y$, because that case is equivalent to Case I by symmetry. Hence we queried $u$ with $y$ and $v$ with $x$, getting answer 2 for both queries. 

If $x$ is not a neighbor of $u$ and $y$ is not a neighbor of $v$ in $T_1$, then $y,u,v,x$ is on a path in this order (but not four consecutive vertices), contradicting Lemma~\ref{lemm1}. Hence, by symmetry, we can assume that $y$ is a neighbor of $v$ in $T_1$, i.e., $uvy$ is a path in $T_1$. 

There are three options. Either $xuvy$ is a path, which is not possible by Lemma~\ref{lemm1}, or $uvyx$ is path, as we wanted, or $x$ is not a neighbor of $u$ and $y$, which is not possible by Lemma~\ref{lemm2}.

We have obtained that either $vuxy$ or $xyvu$ is a path in $T_1$. Hence using Lemma~\ref{lemnew1} it follows that, in the former case, $z=x$, and in the latter case, $z=y$ has the same distance from all other vertices in $T_0$ and $T_1$.
\end{proof}

Now we prove that any two consistent trees are isomorphic. 

\begin{lemma}\label{lem:isom}
    For every pair of consistent trees $T_0, T_1$, there is a graph isomorphism $\varphi \colon V(T_0) \to V(T_1)$. Furthermore, if vertices $x$ and $y$ have distance at least 5 in $T_0$, then they have the same distance in $T_1$.
\end{lemma}

\begin{proof}
Let $u$ be a vertex such that the distance of $u$ and any other vertex $v$ is the same in $T_0$ and $T_1$. Recall $u$ exists by Lemma~\ref{lemnew2}. Let $U_\ell$ denote the set of vertices at distance $\ell$ from $u$ in $T_0$ and $T_1$. We consider $u$ to be the root of trees $T_0$ and $T_1$. In particular, we apply the usual terminology. For a tree $T \in \{T_0, T_1\}$ and a vertex $v\in U_\ell$, the \textit{parent} of $v$ is its neighbor in $U_{\ell-1}$ and its \textit{children} are its neighbors in $U_{\ell+1}$. The \textit{ancestors} of $v$ are the parent of $v$ and parents of its ancestors, the \textit{descendants} of $v$ are the children of $v$ and children of its descendants. The following trivial claim will be important later.

\begin{claim}\label{cla1}
  \textbf{(i)}  If $v\in U_\ell$ and $v'\in U_j$, then $v'$ is a descendant of $v$ in $T \in \{T_0, T_1\}$ if and only if their distance is $j-\ell$ in $T$.

  \textbf{(ii)} If $v\in U_\ell$ and $v'\in U_j$ is a descendant of $v$ in $T \in \{T_0, T_1\}$, then a vertex $v''\in U_i$ is a descendant of $v$ in $T$ if and only if the distance of $v'$ and $v''$ is at most $j+i-2\ell$ in $T$. 
\end{claim}

First, we prove the following.

\begin{claim}\label{cla2}
    If a vertex $v$ has at least three descendants in $T_0$, then the same vertices are its descendants in $T_1$. Similarly, if a vertex $v$ has at least three descendants in $T_1$, then the same vertices are its descendants in $T_0$.
\end{claim}

\begin{proof}[Proof of Claim~\ref{cla2}] 
At least one of the descendants $v'$ of $T_0$ was queried with $v$, thus we know that $v'$ is a descendant of $v$ in $T_1$ as well by Claim~\ref{cla1} (i).

If there is a vertex $v''$ that is a descendant of $v$ in $T_0$ and $v''$ is not a descendant of $v$ in $T_1$, then $v''$ has an ancestor $w\in U_\ell$ in $T_1$ with $w\neq v$. Then $v''$ was not queried with $v$ and $w$, thus $v''$ was queried with $v'$.
Therefore, we get that $v''$ is a descendant of $v$ in $T_1$ as well by Claim~\ref{cla1} (ii), a contradiction.
\end{proof}

We are ready to give an isomorphism $\varphi$ between $T_0$ and $T_1$. First, we define a subtree of $T_0$, where $\varphi$ will be the identity. We call the vertices of this subtree \emph{nice}. We define the nice vertices recursively with respect to the distance from $u$. Let $u$ be nice. Assume that we determined the nice vertices up to distance $\ell-1$ from $u$. We say that $v \in U_\ell$ is nice if the parent of $v$ in $T_0$ (and hence in $T_1$) is nice and the children of $v$ in $T_0$ and $T_1$ are the same. Clearly, the subgraph spanned by the nice vertices is a subtree of $T_0$ and $T_1$, and the identity map between them is an isomorphism of these subtrees. 
The goal is to extend this isomorphism to $V(T_0)$. 

Let $u'\in U_{\ell-1}$ be a nice vertex. It follows that the set of descendants of $u'$ in $T_0$ and $T_1$ are the same. Indeed, let $v$ be one of the closest descendants to $u'$ in $T_0$ that is not a descendant of $u'$ in $T_1$. As $u'$ is nice, $v$ cannot be a child of $u'$. Also, $v$ has distinct parents in $T_0$ and $T_1$, thus was not queried with them. But then $v$ was queried with $u'$, thus, by Claim \ref{cla1} (i), it is a descendant of $u'$ in both $T_0$ and $T_1$, a contradiction. 

Let $U_{\ell}'$ denote the set of not nice vertices in $U_\ell$ that are children of $u'$, i.e., whose descendants are not the same in $T_0$ and $T_1$. Now we will define $\varphi$ on $U'_\ell$ and the descendants of vertices in $U'_\ell$. By the earlier observations, these are the same set of vertices in $T_0$ and $T_1$. By Claim \ref{cla2}, the vertices in $U_{\ell}'$ have at most two descendants in both $T_0$ and $T_1$.
Consider first $u_0\in U_\ell'$ with two children $v,v'$ in $T_0$ such that the parent of $v$ is another vertex $u_1$ in $T_1$. Then $v$ was not queried with $u_0$, nor with $u_1$, thus we know $q(v,v')=2$, hence we know that $v'$ is a child of $u_1$ in $T_1$. By Claim \ref{cla2}, $u_0$ has no other descendants in $T_0$ and $u_1$ has no other descendants in $T_1$. Define $\varphi(v)=v$, $\varphi(v')=v'$ and $\varphi(u_0)=u_1$.
Conversely, if a vertex $u_1 \in U_{\ell}'$ has two children $v$ and $v'$ in $T_1$, then there is a vertex $u_0 \in U_{\ell}'$ with children $v, v'$ in $T_0$, hence we already defined the preimage of $u_1, v, v'$ with respect to $\varphi$ in this step.

Let $U_\ell''$ denote the set of vertices in $U_\ell'$ that do not have two children in $T_0$. As they have at most two descendants, we can partition $U_\ell''$ into $3$ sets $U_{\ell 0}''$, $U_{\ell 1}''$, $U_{\ell 2}''$ with $U_{\ell 0}''$ consisting of vertices with no children, $U_{\ell 1}''$ being the set of vertices that have one child with no children, and $U_{\ell 2}''$ having vertices with one child who has one child with no children $T_0$. Similarly, let $V_\ell''$ denote the set of vertices in $U_\ell'$ that do not have two children in $T_1$. We can partition $V_\ell''$ into $3$ sets $V_{\ell 0}''$, $V_{\ell 1}''$, $V_{\ell 2}''$ with $V_{\ell 0}''$ consisting of vertices with no children, $V_{\ell 1}''$ being the set of vertices that have one child with no children, and $V_{\ell 2}''$ having vertices with one child who has one child with no children in $T_1$. 
We have $|U_{\ell i}''|=|V_{\ell i}''|$ for $i=0,1,2$, as the number of descendants of $u'$ in $U_{\ell+1}$ and $U_{\ell+2}$ is the same in $T_0$ and $T_1$.
Define $\varphi$ on $U_\ell''$ by taking arbitrary bijections between vertices in $U_{\ell i}''$ and $V_{\ell i}''$ for $i=1,2,3$. Finally, if a vertex $x$ is the child of some vertex $z \in U_\ell''$, then let $\varphi(x)$ be the child of $\varphi(z)$, and similarly, if $x$ is the child of the child of some vertex $z \in U_\ell''$, then let $\varphi(x)$ be the child of the child of $\varphi(z)$.

Doing this for all nice vertices $u'$ defines the mapping $\varphi$ on every vertex of $V(T_0)$, and $\varphi$ is clearly an isomorphism by definition.

From the above definitions and observations, it is not hard to prove the second part of the lemma. Consider vertices $x$ and $y$ with distance at least 5 in $T_0$. Clearly, if $x$ or $y$ is nice, then their distance is the same in $T_0$ and $T_1$. Let $x'$ and $y'$ be the closest nice ancestors of $x$ and $y$, respectively. By the nice property, these are the same vertices in $T_0$ and $T_1$. It is also clear that in the case of $x' \neq y'$, the distance of $x$ and $y$ is the same in $T_0$ and $T_1$. Hence, assume that $x'=y'$ and let $x' \in U_{\ell-1}$. Let $x \in U_{\ell+a}$ and $y \in U_{\ell+b}$. As by Claim \ref{cla2}, children of $x'$ that are ancestors of either $x$ or $y$ have at most two descendants, we know that $a,b \leq 2$. 

Furthermore, as the distance of $x,y$  is at least 5 in $T_0$, $a+b \geq 3$ and $x$ and $y$ has $x'$ as the closest common ancestor in $T_0$. By Claim \ref{cla2}, there cannot be a common ancestor $u \in U_{l} \cup U_{l+1} \cup U_{l+2}$ with at least 3 descendants in $T_1$.

Assume first that $a=b=2$. Then $d(x,y)$ in $T_1$ could be either $2,4$, or $6$. If either $d(x,y)=2$ or $d(x,y)=4$, the common grandparent $u$ of $x$ and $y$ in $T_1$ has at least $3$ descendants, a contradiction by the previous observation.

Now, assume by symmetry that $a=2, b=1$. We need to show that $d(x,y)$ in $T_1$ cannot be $1$ or $3$. If $d(x,y)=3$, then their closest common ancestor in $T_1$ has at least $3$ descendants, which is again a contradiction. In the remaining case, $y$ is the parent of $x$ in $T_1$. Let $w_1$ be the parent of $x$, $w_2$ be the grandparent of $x$, and $w_3$ be the parent of $y$ in $T_0$. As we have two vertices, $y$ and $w_1$, whose distance from $x$ differs in $T_0$ and $T_1$, we must have that $w_2$ is the grandparent of $x$ in $T_1$ as well, hence it is the parent of $y$. Then the distances of $y$ from $x$, $w_2$, and $w_3$ are all different in $T_0$ and $T_1$, a contradiction.
\end{proof}

It remains to show that we can always find two vertices with the largest distance.

\begin{lemma}\label{lem:diam4}
    If the diameter of any consistent tree $T_0$ is at least $4$, then there is a vertex pair which has the same distance $j$ in all consistent trees for some $j \geq 4$.
\end{lemma}
\begin{proof} Note that if the answer to any query is at least $4$, then we are done. 
Take two vertices $u,v$ with the largest distance in a consistent tree $T_0$. If their distance is at least $5$, then by Lemma \ref{lem:isom}, their distance is the same in each consistent tree. Therefore, we may assume that their distance is $4$ in $T_0$. If their distance is less than $4$ in some consistent tree $T_1$, then by Lemma \ref{parity}, it can only be $2$.

Suppose that $uw_1w_2w_3v$ is a path in $T_0$ while $uxv$ is a path in $T_1$, where $x$ is not necessarily distinct from $w_1, w_2$, or $w_3$. Notice that $u,v$ were not queried, and at least one of $u$ and $v$ was not queried with $x$. As in the proof of Lemma~\ref{null}, $n \geq 13$ implies that there exists a vertex $y \notin\{u,w_1,w_2,w_3,v\}$ that was queried with all of these five vertices.

As $T_1$ is consistent, there are three options on the  difference of the distances from $y$ to $u$ and $v$.

CASE I: $q(y,u)+2=q(y,v)$. Then $y$ must be closest to $w_1$ in $T_0$ among $\{u,w_1,w_2,w_3,v\}$, therefore $q(y,v) \geq 4$, so we queried a distance which is at least $4$.

CASE II: $q(y,u)=q(y,v)+2$. This follows similarly as the previous case.

CASE III: 
$q(y,u)=q(y,v)$. Then $y$ must be closest to $w_2$ in $T_0$ among $\{u,w_1,w_2,w_3,v\}$, in particular, $q(y,w_2)=1$, $q(y,w_1)=q(y,w_2)=2$ and $q(y,u)=q(y,v)=3$, otherwise we had a distance at least 4 query, and we are done. It follows that $y$ is closest to $x$ in $T_1$ and their distance is 2. Suppose that $x\neq w_1$. Then the distance of $u$ and $w_1$ is different in the two trees, thus $u$ was not queried with $w_1$ nor $v$, therefore $q(u,w_3)=3$, and $x \neq w_3$. By symmetry, this implies that $x$ is distinct from $w_1$ and $w_3$, $u$ was not queried with $w_1$, and $v$ was not queried with $w_3$. Recalling that $u,v$ were not queried, and $x$ was not queried with $u$ or $v$, we reach a contradiction as the degree of $u$ or $v$ would be at most $n-4$.    
\end{proof}

Finally, we prove Theorem~\ref{nemad}.

\begin{proof}[Proof of Theorem \ref{nemad}]

The lower bound was proved in Lemma~\ref{lemi}. 

For the upper bound, the isomorphism part follows by Lemma~\ref{lem:isom}.
It remains to show two vertices with the largest distance, as it does not follow from knowing the graph up to isomorphism. However, by the second part of Lemma~\ref{lem:isom}, if two vertices $x$ and $y$ have distance at least 5 in a consistent tree, then their distance is the same in all consistent trees. 
Therefore, if the diameter is at least $5$, we can show two vertices whose distance is the diameter. Otherwise, we know that the diameter is at most $4$, and if there is a consistent tree with diameter $4$, then by Lemma \ref{lem:diam4}, we can find a pair of vertices with distance $4$. 

The only remaining case is when all consistent trees have diameter at most $3$. In this case, observe that if a tree on at least 7 vertices is not a star, then it has a vertex that has distance 3 from at least 3 vertices. Therefore, at least one query was answered 3. Hence the largest of the known distances is the largest distance. For stars, all the answers are 1 or 2, thus we know we have a star. If $n\ge 5$, at least one answer was 2, and that is the largest distance.

\end{proof}

\smallskip
\textbf{Acknowledgement}: Research was carried out at the Combinatorial Search Theory Seminar of the Alfr\'ed R\'enyi Institute of Mathematics. 

\smallskip
\textbf{Funding}: Research was supported by the National Research, Development and Innovation Office - NKFIH under the grant KKP-133819. The second author was supported by the EKÖP-24 University Excellence Scholarship Program of the
Ministry for Culture and Innovation from the source of the National
Research, Development and Innovation Fund.
The fifth author was supported by the Thematic Excellence Program TKP2021-NKTA-62 of the National Research, Development and Innovation Office.

\end{document}